\documentclass[traditabstract]{aa}
\usepackage{natbib,graphicx,hyperref} %,amsmath,amssymb,graphicx,hyperref}
\usepackage[varg]{txfonts}
\begin{document}

\title{Origin of the hemispheric asymmetry of solar activity}

\author{M.~Sch{\"u}ssler
\inst{1}
\and
R.~H.~Cameron\inst{1}
}

\institute{Max-Planck-Institut fÃ{\"u}r Sonnensystemforschung
Justus-von-Liebig-Weg 3, 37077 G{\"o}ttingen\\
\email{cameron@mps.mpg.de}
}

\date{Received ; accepted }

% \abstract{}{}{}{}{}
% 5 {} token are mandatory
% context heading (optional)
% {} leave it empty if necessary
% aims heading (mandatory)
% methods heading (mandatory)
% results heading (mandatory)
% conclusions heading (optional), leave it empty if necessary

\abstract {The frequency spectrum of the hemispheric asymmetry of
  solar activity shows enhanced power for the period ranges around 8.5
  years and between 30 and 50 years. This can be understood as the
  sum and beat periods of the superposition of two dynamo modes: a
  dipolar mode with a (magnetic) period of about 22 years and a
  quadrupolar mode with a period between 13 and 15 years. An updated
  Babcock-Leighton-type dynamo model with weak driving as indicated by
  stellar observations shows an excited dipole mode and a damped
  quadrupole mode in the correct range of periods. Random excitation
  of the quadrupole by stochastic fluctuations of the source term for
  the poloidal field leads to a time evolution of activity and
  asymmetry that is consistent with the observational results.}

\keywords{Sun: activity -- Sun: magnetic fields -- Dynamo}

\titlerunning{Solar hemispheric asymmetry}
\maketitle
%
%-------------------------------------------------------------------

\section{Introduction}
\label{sec:introduction}

The various manifestations of solar magnetic activity, such as
sunspots, prominences, and flares, typically are distributed unevenly
between the northern and southern hemisphere of the Sun \citep[cf.][and
references therein]{Norton:etal:2014, Hathaway:2015, Deng:etal:2016}.
Normally, this hemispheric asymmetry does not exceed a level of about
20\% \citep{Norton:Gallagher:2010}. However, during the Maunder minimum
in the second half of the 17th century nearly all of the few sunspots
observed during this time appeared in the southern hemisphere
\citep{Ribes_Nesme-Ribes:1993, Vaquero:etal:2015}. Various studies
demonstrated that the asymmetry has systematic components that cannot be
explained by random fluctuations of flux emergence alone
\citep[e.g.][]{Carbonell:etal:1993, Carbonell:etal:2007,
Deng:etal:2016}. It has been suggested that there is also a systematic
variation of the phase shift of the activity cycle between the
hemispheres \citep[e.g.][]{Zolotova:etal:2010, Norton:Gallagher:2010,
Murakoezy:Ludmany:2012, McIntosh:etal:2013}.

A number of studies investigate the hemispheric asymmetry by way of
frequency analysis \citep[e.g.][and references
therein]{Deng:etal:2016}. \citet{Ballester:etal:2005} demonstrated that
the commonly used normalised asymmetry parameter, $(N-S)/(N+S)$,
where $N$ and $S$ represent the quantity under consideration in the
northern and southern hemisphere, respectively, is not a sensible choice
for this kind of analysis: the denominator introduces a contamination of
the power spectrum by the strong 11-year
periodicity. \citet{Ballester:etal:2005} instead carry out a frequency
analysis of the un-normalised asymmetry, $A_{\rm N}-A_{\rm S}$, of
the monthly hemispheric sunspot areas between 1874 and 2004. They use
the dataset compiled by D. Hathaway, based upon the Greenwich
Photoheliographic Results and the USAF/SOON data, and find three
significant periods with false-alarm probabilities below 0.5\%: 43.25,
8.65, and 1.44 years. Very similar periods (among others) are also
found by \citet{Knaack:etal:2004}, who use the same dataset, while
\citet{Deng:etal:2016} reported periods of 51.3 and 8.7 years. Studying
the hemispheric asymmetry of filaments between 1919 and 1989,
\citet{Duchlev:Dermendjiev:1996} find periods of 35 and 8.75 years,
although they considered only the former to be statistically
significant. On the other hand, \citet{Chang:2009} suggests that only
the period around nine years in the asymmetry of sunspot areas is
significant while other periodicities may not. Since the length of the
various data series does not exceed about 150 years, the frequency
resolution for the longer periods is rather low. Thus we may conclude
from these studies that there is evidence for a short period
(around 9 years) and a long period (between 35 and 50 years) in the data
for the un-normalised asymmetry, while the 11-year cycle does not
significantly appear.

In this paper, we show that the periods found in the observational data
for the absolute hemispheric asymmetry occur naturally as the beat
period and the sum period of a mixed-mode dynamo solution comprised of a
dipole mode with a (magnetic) period of about 22 years and a quadrupole
mode with a period between 13 and 15~years. We also find that periods in
this range are reproduced by the updated Babcock-Leighton dynamo model
of \citet{Cameron:Schuessler:2017a} in the case of weak dynamo driving as
suggested by stellar observations. While the dipole mode is permanently
excited, the quadrupole is subcritical and only occasionally kicks in
through random fluctuations of the poloidal source term.

This paper is structured as follows. In Sect.~\ref{sec:model}, we
present a simple model of superposed harmonic oscillations to illustrate
the origin of the various periodicities. From the observed periods of
the hemispheric asymmetry, we determine the periods of the antisymmetric
(dipole) mode and the symmetric (quadrupole) mode. Sect.~\ref{sec:BL}
gives the corresponding results obtained with the updated
Babcock-Leighton model. We summarise our conclusions in
Sect.~\ref{sec:conclusions}.

\section{Hemispheric aymmetry by superposition of symmetric and
antisymmetric modes}
\label{sec:model}
As a simple illustration of the possible origin of the various periods
detected in the sunspot area data (full disk, hemispheric, and
asymmetry), we have considered the superposition of two harmonic
oscillations with different frequencies. They are taken to represent two
dynamo modes for the toroidal field, $B_\phi$: one mode is antisymmetric
with respect to the equator (dipole parity, frequency $\omega_{D}$), the
other mode is symmetric (quadrupole parity, frequency
$\omega_{Q}$). Since we are only interested in the frequencies resulting
from the superposition, we set the amplitudes of the modes to be equal
and normalise them to unity. Taking the activity index to be
proportional to $B^2_\phi$, that is, the square of the superposed modes, we
have for the indices in the northern hemisphere, $A_{\rm N}$, and in the
southern hemisphere, $A_{\rm S}$, respectively:
\begin{eqnarray}
A_{\rm N} &=& \left[ \sin(\omega_D t) + \sin(\omega_Q t) \right]^2 , \;
{\rm and}
\nonumber \\
A_{\rm S} &=& \left[ -\sin(\omega_D t) + \sin(\omega_Q t) \right]^2 .
\label{eq:ANS}
\end{eqnarray}
The indices for the full disk and for the (absolute) asymmetry are then
given by the sum and the difference, respectively, of the hemispheric
signals, viz.
\begin{eqnarray}
A_{\rm N} + A_{\rm S} &=& 2\sin^2(\omega_D t) + 2\sin^2(\omega_Q t)
\nonumber \\
A_{\rm N} - A_{\rm S} &=& 4\sin(\omega_D t) \sin(\omega_Q t)
\nonumber \\
&=& 2\cos[(\omega_Q-\omega_D)t] - 2\cos[(\omega_Q + \omega_D)t] \,,
\label{eq:sum_dif}
\end{eqnarray}
where $\omega_Q - \omega_D\equiv\omega_b$ is the beat frequency and
$\omega_Q + \omega_D\equiv \omega_s $ is the sum
frequency. It is clear from Eqs.~(\ref{eq:ANS}) and (\ref{eq:sum_dif})
that the frequencies appearing in the various quantities are different:
while only the double frequencies, $2\omega_D$ and $2\omega_Q$, show up
in the full-disk index, the absolute asymmetry is governed solely by the
beat and the sum frequencies. The hemispheric indices are affected by
all four of these frequencies. In terms of periods, we have
\begin{equation}
P_b = \frac{P_D P_Q}{P_D - P_Q}
\label{eq:beat_period}
\end{equation}
for the beat period and
\begin{equation}
P_s = \frac{P_D P_Q}{P_D + P_Q}
\label{eq:sum_period}
\end{equation}
for the period corresponding to the sum frequency, where
$P_D=2\pi/\omega_D$ and $P_Q=2\pi/\omega_Q$, respectively, are the
periods of the dipole and quadrupole dynamo modes.

Consistent with the expectation from this simple model, the analysis of
the absolute asymmetry of hemispheric sunspot areas by
\citet{Ballester:etal:2005} results in two dominant periods,
$43.25\,$years and $8.65\,$years, while the 11-year cycle period
(dominated by the dipole) does not appear. Tentatively identifying
these two observed periods with the beat and sum periods, we can use
Eqs.~(\ref{eq:beat_period}) and (\ref{eq:sum_period}) determine the
dipole period, $P_D$, and the quadrupole period, $P_Q$. With $P_b =
43.25\,$years and $P_s=8.65\,$years we obtain $P_D=21.6\,$years and
$P_Q=14.4\,$years. If we take into account the limited frequency
resolution and assume a range between 30 years and 50 years for the for
the longer (beat) period, we find $P_D \approx 21...24\,$years and $P_Q
\approx 13...15\,$years. The obtained dipole period is consistent with
the 11-year activity cycle. This suggests that our simple model is a
viable representation of the asymmetry data, suggesting the presence of
a solar quadrupole dynamo mode with a period between 13 and $15\,$years.
Incidentally, \citet{Munoz:etal:2013} show that considering both the
dipole and quadrupole moments of the poloidal field during cycle minima
improves the predictive power for the amplitude of the subsequent cycle
\citep[see also][]{Goel:Choudhuri:2009}.

\begin{figure*}
\centering
\includegraphics[width=\textwidth]{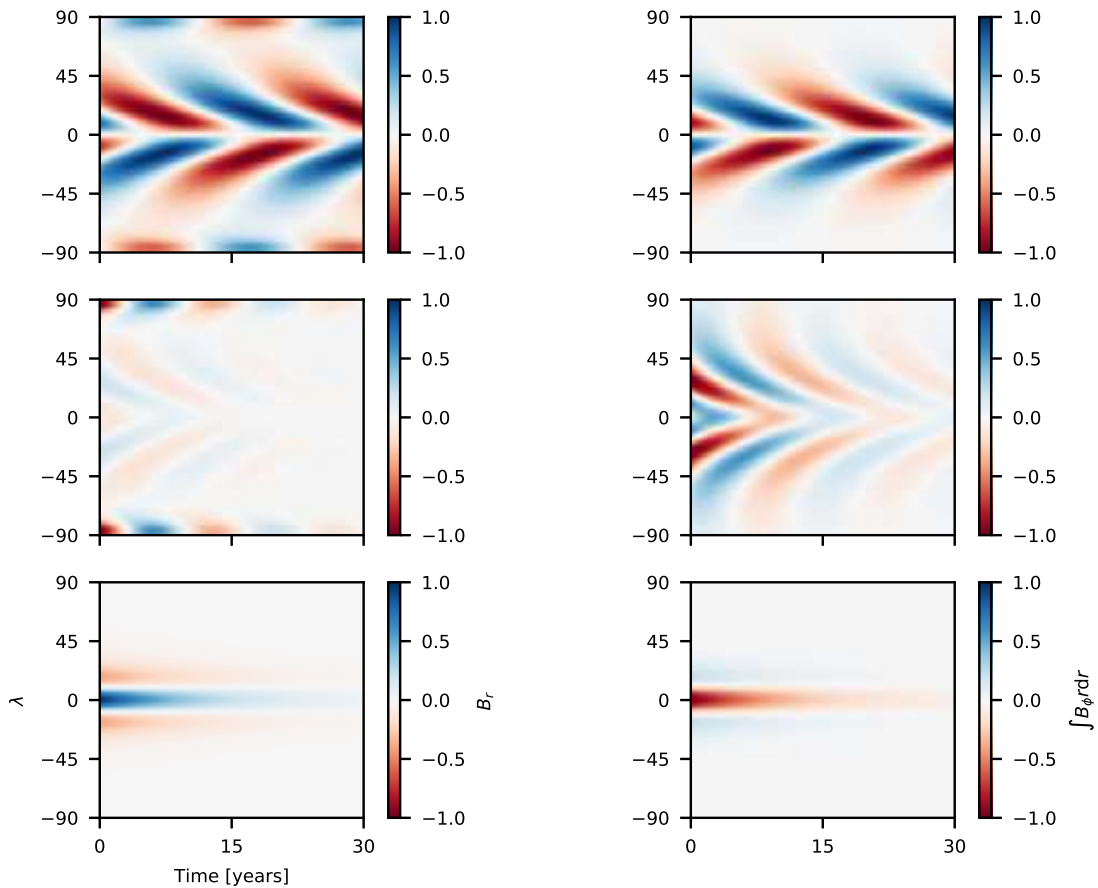}
\caption{Spatio-temporal structure of the first linear eigenmodes
obtained with the updated Babcock-Leighton model
\citep{Cameron:Schuessler:2017a}. Shown are latitude-time
diagrams of the azimuthally averaged radial surface field
(left panels) and the radially integrated toroidal flux (right
panels). The three modes shown are an excited oscillatory
dipolar mode (top panels), a damped oscillatory quadrupolar mode
(middle panels) and a damped stationary quadrupolar mode
(bottom panels). The quantities are normalised
to their individual maxima in all cases.}
\label{fig:modes}
\end{figure*}

\section{Dynamo model}
\label{sec:BL}

\begin{figure*}
\centering
\includegraphics[width=\textwidth]{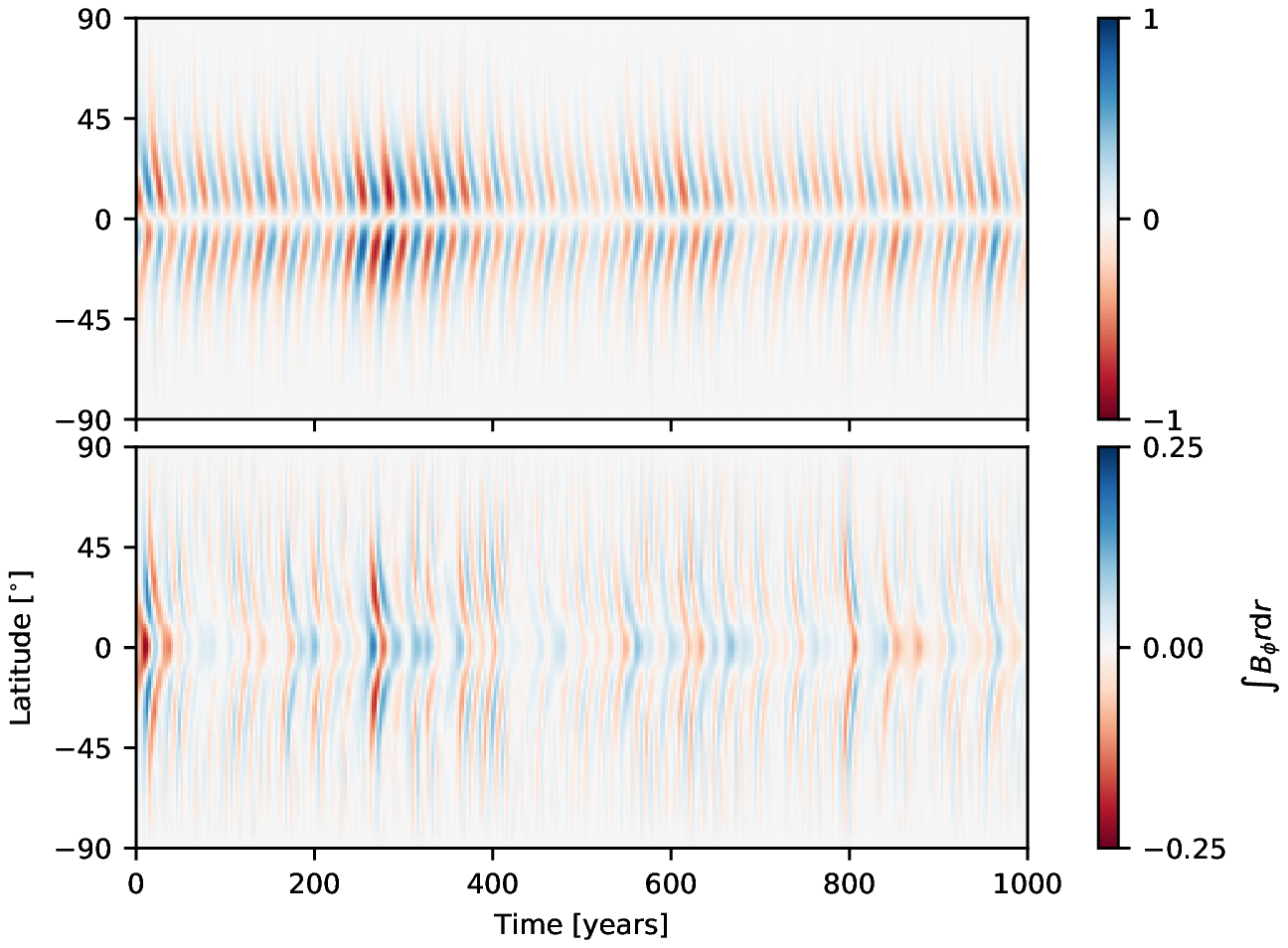}
\caption{Non-linear case with random fluctuations of the poloidal
source term. Shown are time-latitude diagrams of the
antisymmetric (dipolar, upper panel) and symmetric
(quadrupolar, lower panel) components of the radially
integrated toroidal magnetic flux. The linearly damped
oscillatory and stationary quadrupolar modes are occasionally
excited.}
\label{fig:longterm}
\end{figure*}

Hemispheric asymmetry in dynamo models has been studied by various
authors \citep[reviewed by][]{Norton:etal:2014, Brun:etal:2015}. There
are two main approaches that have been followed: non-linear effects and
stochastic fluctuations. Non-linearity in the dynamo equations can lead
to the coupling of symmetric (even) and antisymmetric (odd) modes,
strong hemispheric asymmetry, and the occurence of extended `grand
minima' \citep[e.g.][]{Kleeorin:Ruzmaikin:1984, Tobias:1997,
Brooke:etal:1998, Hotta:Yokoyama:2010, Weiss:Tobias:2016}.
Alternatively, stochastic fluctuations of model ingredients (such as
mean-field $\alpha$-effect or meridional flow speed) can also lead to
hemispheric asymmetry as well as to the (temporal) excitation of higher
eigenmodes and mixed-mode solutions \citep[e.g.][]{Hoyng:etal:1994,
Olemskoy:Kitchatinov:2013, Belucz:Dikpati:2013,
Passos:etal:2014}. Combinations of both effects have been studied as
well \citep[e.g.][]{Moss:etal:1992, Schmitt:etal:1996,
Mininni:Gomez:2002, Charbonneau:2007, Moss:Sokoloff:2017}.  For example,
\citet{Sokoloff:etal:2010} and \citet{Usoskin:etal:2009} find that
random fluctuations of the dynamo excitation in a simple Parker-type
dynamo wave model can lead to substantial mixing between dipole and
quadrupole modes, particularly so during episodes of low dynamo
amplitude akin to grand minima of solar activity.  Global 3D-MHD
simulations exhibit features that are similar to the results provided by
the more idealised approaches \citep{Norton:etal:2014,
Brun:etal:2015,Kaepylae:etal:2016}.

Dominance of the dipole mode and relatively weak hemispheric asymmetry
can be provided by sufficiently strong hemispheric coupling via
turbulent diffusion, cross-equator flows, or cross-equator cancellation
of toroidal flux \citep{Norton:etal:2014, Cameron:Schuessler:2016}.
Moreover, observational gyrochronology of solar-like stars
\citep{vanSaders:etal:2016, Metcalfe:etal:2016} indicates weak
excitation of the solar dynamo. In this case, there is the possibility
that only the lowest (dipole) dynamo mode is excited while the
quadrupole mode is linearly damped.

As an illustration for the stochastic excitation of a mixed-mode dynamo
solution that is consistent with the observed features of the
hemispheric asymmetry, we show results from the updated Babcock-Leighton
model of \citet{Cameron:Schuessler:2017a} with stochastic fluctuations
of the poloidal field source \citep[see Sect. 3
in][]{Cameron:Schuessler:2017b}. The parameters for this case were
chosen according to the following observational constraints: (a) excited
dipole mode with a period of about 22 years, (b) phase difference
between polar radial field and subsurface toroidal flux of about
$90^\circ$, (c) no linearly excited quadrupolar mode. In such a case,
linearly damped quadrupole modes can be stochastically excited by the
fluctuations of the poloidal field source, so that a mixed-mode solution
occasionally develops. The parameters for the case discussed here were:
$\eta_{R_{\odot}}=150$~km$^2\cdot$s$^{-1}$ and
$\eta_0=50$~km$^2\cdot$s$^{-1}$ for the turbulent magnetic diffusivities
in the near-surface layers and in the bulk of the convection zone,
respectively; $\alpha_0=1.3$~m$\cdot$s$^{-1}$ for the average poloidal
source level, and $\sigma^* = 0.046$ for the fluctuation level.  The
source fluctuations are local in latitude (in steps of 1 degree) and in
time (in steps of 1 day), governed by a Wiener process with a variance
of 1 radian$^{-1}$ after 11 years. This corresponds to a RMS fluctuation
of the source term of about 5\% (integrated over 11 years and one
radian).  We used the near-surface meridional flow as determined by
\citet{Hathaway:Rightmire:2011} and $V_0=1.7$~m$\cdot$s$^{-1}$ for the
amplitude of the equatorward meridional return flow affecting the
toroidal flux in the convection zone. The critical integrated flux for
the cut-off non-linearity in the poloidal source term was $10^{24}\,$Mx.

\begin{figure}
\centering
\includegraphics[width=0.43\textwidth]{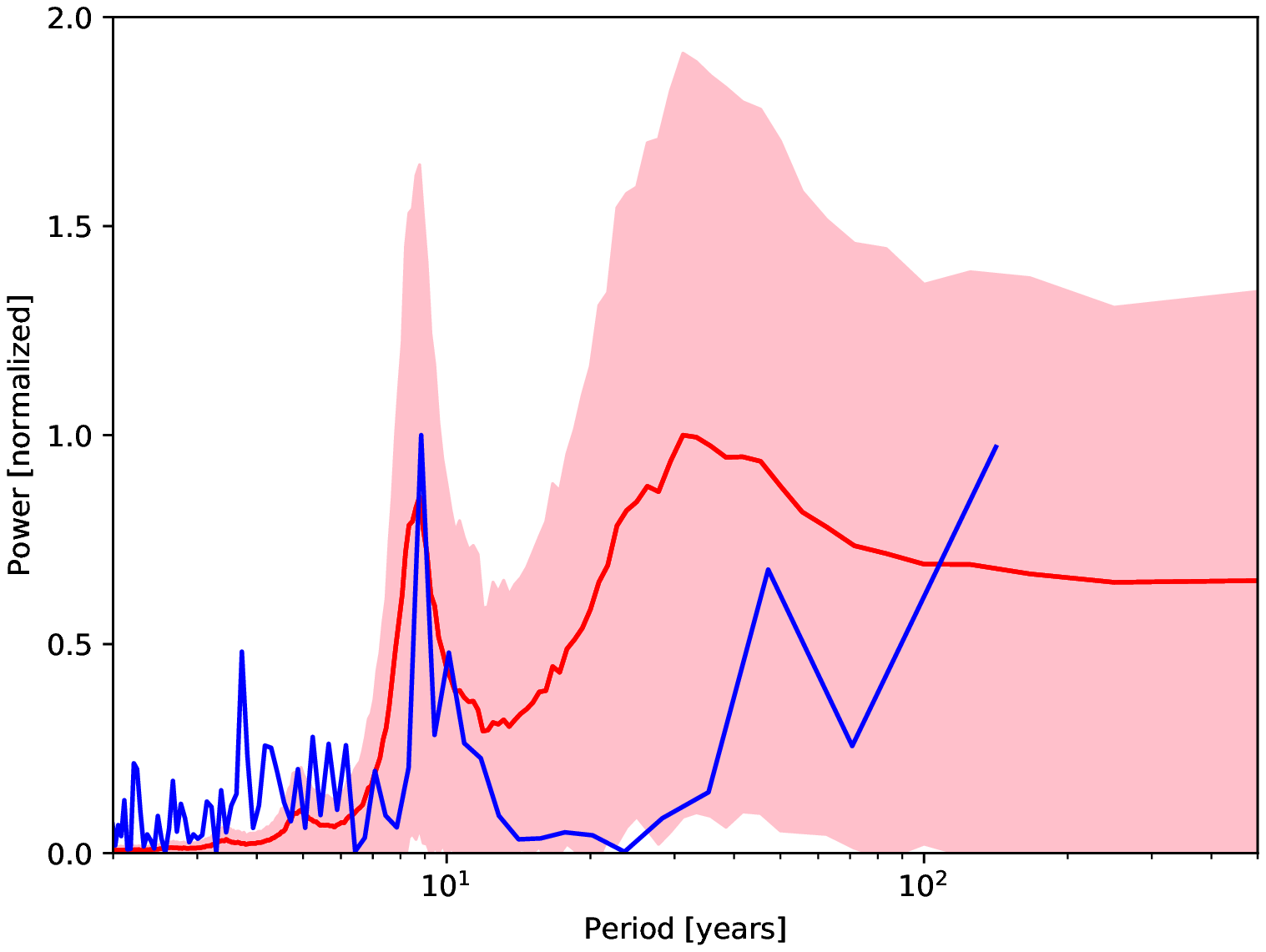}
\caption{Power spectra of the absolute hemispheric asymmetry. The blue
curve corresponds to the absolute asymmetry of the sunspot
areas from the Greenwich/SOON data. The red curve gives the
average power spectrum for the difference of the unsigned
toroidal fluxes integrated between the equator and $\pm
30/40^\circ$ latitude). The latter is based on 1000
realizations of 500 year each from the Babcock-Leighton model
with source fluctuations \citep{Cameron:Schuessler:2017b}. The
pink shading indicates $\pm$1 standard deviation from the
mean.}
\label{fig:power_asym}
\end{figure}

Linear analysis shows that for these parameters we have an excited
oscillating antisymmetric (dipolar) mode with a period of 22.4 years
($\alpha/\alpha_{\rm crit}=1.06$) and a damped symmetric (quadrupolar)
mode with a period of 13.5 years and a damping time of 13 years
($\alpha/\alpha_{\rm crit}=0.12$).  In addition, there is a symmetric
stationary mode with a damping time of about 10
years. Fig.~\ref{fig:modes} shows the spatio-temporal structure of these
linear modes.  Fig.~\ref{fig:longterm} shows the symmetric and
antisymmetric components of a simulation of the non-linear case with
fluctuations. The quadrupolar (symmetric) modes are occasionally excited
owing to the random fluctuations of the poloidal source term. In
addition, there are variations of the dynamo amplitude.  The ratio of
the RMS values of the toroidal field variable between the quadrupole and
the dipole modes is about 0.18.  Owing to the non-linearity, the
oscillation periods become somewhat variable and their mean values
differ from their linear counterparts: 20.8 years for the dipole and
14.7 years for the oscillatory quadrupole. According to
Eqs.~(\ref{eq:beat_period}) and (\ref{eq:sum_period}), the linear
periods lead to a beat period $P_b=34\,$years and a sum period
$P_s=8.4\,$years while the non-linear periods give $P_b=50.1\,$years and
$P_s=8.6\,$years. Hence, the beat period is much more sensitive to
period variations than the sum period. We therefore expect periodic
signals in the absolute asymmetry (for instance of the toroidal flux
taken as an activity indicator) around 8.5~years ($P_s$) and in the
range 30-50 years ($P_b$). Such signals can in fact be seen in
Fig.~\ref{fig:power_asym}, which shows power spectra of the absolute
hemispheric asymmetry of the hemispheric toroidal flux integrated
between $0^\circ$ and $\pm 40^\circ$ latitude (red curve) from the
Babcock-Leighton model in comparison with that of the observed sunspot
area (blue curve). This is consistent with the expectation from the
simple model described in Sec.~\ref{sec:model}. The widths of the peaks
results from the variability of the (non-linear) periods, the damping of
the quadrupole mode, as well as from realisation noise. The dynamo model
serves only as an illustration of the proposed mechanism behind the
hemispheric asymmetry; we have made no attempt to fine-tune the
parameters in order to have a precise agreement of the model result with
the observed peak at 43 years. It suffices to state that the updated
Babcock-Leighton model with low excitation and stochastic fluctuations
of the poloidal source yields results that are consistent with the
observed features.

\begin{figure}
\centering
\includegraphics[width=0.47\textwidth]{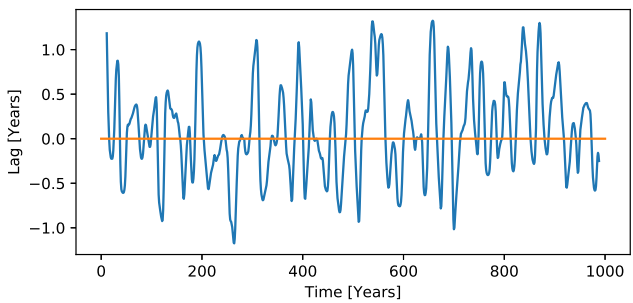}
\caption{Phase lag (time shift) between the toroidal fluxes in the
northern and southern hemispheres for a 1000-yr stretch of a simulation
on the basis of the updated Babcock-Leighton model. The lag was
determined by using the cross-covariance between segments with a length
of 20 years.}
\label{fig:lag}
\end{figure}

\begin{figure}
\centering
\includegraphics[width=0.43\textwidth]{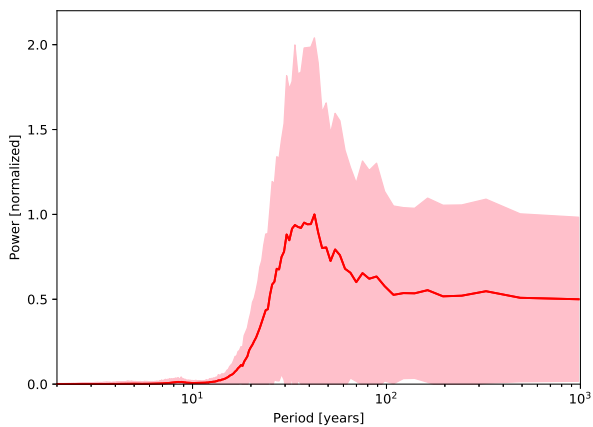}
\caption{Average power spectrum of the time shift between the
hemispheres from 20 realizations of the updated Babcock-Leighton dynamo
of 10,000 year length each. The shading indicates $\pm 1$ standard
deviations.}
\label{fig:power_lag}
\end{figure}

We may also consider the phase lag between the hemispheres. From the
sunspot record, some authors suggest a periodicity of the phase
lag of eight cycles, that is, about 90 years
\citep[e.g.][]{Zolotova:etal:2010, Norton:Gallagher:2010,
Murakoezy:Ludmany:2012}, or even twelve cycles
\citep[e.g.][]{Zhang:Feng:2015}. Some caution seems to be in order in
view of the fact that the time series used are not much longer than the
inferred periods. Using a cross-correlation method,
\citet{McIntosh:etal:2013} find long-term variations with the
hemispheres alternating in phase shift for intervals between 30 and 60
years since 1874. On the other hand, on the basis of the simple model in
Sec.~\ref{sec:model}, we expect a periodicity of the phase lag with the
beat period, $P_b$. We have applied a method similar to that used by
\citet{McIntosh:etal:2013} to the results from the updated
Babcock-Leighton model, simulated with the same parameters as described
above. Using the hemispheric toroidal flux integrated from the equator
to $\pm 40^\circ$ latitude, we considered 20-year segments from 50
simulations covering 10,000 years each. Removing the mean of the signals
and applying a Hann window to the segments, we then calculated the
cross-covariance between the northern and southern hemisphere signals
and determined the time lag between the hemispheres by considering the
maximum of the cross-covariance. A typical example for the temporal
variation of the lag is shown in Fig.~\ref{fig:lag}. It suggest a
modulation with a period around 30 years. This is confirmed by the power
spectrum shown in Fig.~\ref{fig:power_lag}. The curve gives the mean
power spectrum for 1000 realizations of 500 years length each while the
shading indicates $\pm 1$ standard deviation. Comparison with
Fig.~\ref{fig:power_asym} shows that the main power appears in the range
of the beat period of 30-50 years, but there is considerable power at
longer periods as well, indicating longer-term modulations. This is also
obvious from Fig.~\ref{fig:lag}, where the sign of the lag often remains
the same for intervals exceeding the beat period.

\section{Conclusions}
\label{sec:conclusions}
We have shown that the observed power spectrum of the absolute
hemispheric asymmetry of solar activity can be naturally explained by
the superposition of an excited dipolar mode (toroidal field
antisymmetric with respect to the equator) with a magnetic period of
about 22 years and a linearly damped, but randomly excited quadrupolar
mode (toroidal field symmetric with respect to the equator) with a
period between 13 and 15 years. The updated Babcock-Leighton dynamo
model of \citet{Cameron:Schuessler:2017a} with weak excitation
reproduces these conditions and yields a time evolution of the magnetic
field and its asymmetry that is consistent with the observations.

\newpage

\bibliographystyle{aa}
\bibliography{Asym}

\end{document}